\documentclass{desyproc}

\newcommand{\ba}{\begin{eqnarray}}
\newcommand{\ea}{\end{eqnarray}}
\newcommand{\beqs}{\begin{eqnarray}}
\newcommand{\eeqs}{\end{eqnarray}}

\begin{document}
\title{Models of the hadron structure and
Data of the TOTEM Collaboration}

\author{{\slshape Oleg Selyugin$^1$}\\[1ex]
$^1$BLTPh, JINR,  Dubna, Russia}

\contribID{smith\_Selyugin}


\acronym{EDS'09} 

\maketitle

\begin{abstract}
  The region of
     the small and large momentum transfer is examined from a
     view point of the contribution of the different parts of
     the scattering amplitude, soft and hard pomeron, and odderon
     contribution. The new model taking into account the
     different moments of the General Parton Distribution of
     the hadron is presented.
  The comparison with the preliminary data of the TOTEM Collaboration at an energy of $7$ TeV is made.
\end{abstract}


There are 
 many different models for the description of hadron elastic
 scattering at small angles \cite{Rev-LHC}. 
   In the Chow-Yang model \cite{CY-65}
    it was assumed that the hadron interaction  was   proportional to the overlapping of the matter distribution of the hadrons, and  Wu and Yang
      \cite{WY-68}
    suggested that the matter distribution  was proportional to the charge distribution of the hadron.
 Then many models  used  the electromagnetic form factors of the hadron
but, in most part, they changed its form to describe the experimental data,
 as was made in  the  
   Bourrely-Soffer-Wu   model \cite{BSW}.

 The differential cross
  sections of nucleon-nucleon elastic scattering  can be written as the sum of different  five
  helicity  amplitudes.
   The total helicity amplitudes can be written as $\Phi_{i}(s,t) =
  F^{h}_{i}(s,t)+F^{\rm em}_{i}(s,t) e^{\varphi(s,t)} $\,, where
 $F^{h}_{i}(s,t) $ comes from the strong interactions,
 $F^{\rm em}_{i}(s,t) $ from the electromagnetic interactions and
 $\varphi(s,t) $
 is the interference phase factor between the electromagnetic and strong
 interactions \cite{selmp1}. 

      Our model is based on the representation that at high energies a hadron interaction in the nonperturbative regime
      is determined by the reggenized-gluon exchange. The cross-even part of this amplitude can have $2$ nonperturbative parts, possible standard pomeron - $P_{2np}$ and cross-even part of the 3-non-perturbative gluons - $P_{3np}$.
      The interaction of these two objects is proportional to two different form factors of the hadron.
      This is the main assumption of the model.
      The second important assumption is that we chose the slope of the second term
       $4$ times smaller than the slope of the first term, by  analogy with the two pomeron cut.
      Both terms have the same intercept.

      The form factors are determined by the General parton distributions of the hadron (GPDs).
      The first form factor corresponding to the first momentum of GPDs is the standard electromagnetic
      form factor - $G(t)$. The second form factor, determined by the second momentum of GPDs -$A(t)$,
      corresponds to the matter  distribution of the nucleon \cite{Pagels,Polyakov1,Polyakov2}.
      The parameters and $t$-dependence of the GPDs are determined by the standard parton distribution
      functions, so by the experimental data on the deep inelastic scattering and by the experimental data
      for the electromagnetic form factors (see \cite{ST-PRDGPD}).

      Hence, the Born term of the elastic hadron amplitude can be written as
 \begin{eqnarray}
 F_{h}^{Born}(s,t)  =  h_1 \ G^{2}(t) \ F_{a}(s,t) \ (1+r_1/\hat{s}^{0.5}) +  h_{2} \  A^{2}(t) \ (F_{b}(s,t) \ (1+r_2/\hat{s}^{0.5})+ F_{odd}(s,t))
\end{eqnarray}
  where $F_{a}(s,t)$ and $F_{b}(s,t)$  has the standard Regge form 
  \begin{eqnarray}
 F_{a}(s,t) \ = \hat{s}^{\epsilon_1} \ e^{B(s) \ t}; \ \ \
 F_{b}(s,t) \ = \hat{s}^{\epsilon_1} \ e^{B(s)/4 \ t};   \ \ \  F_{odd}(s,t)=\frac{i t}{1/r_{0}^2-t}  \hat{s}^{\epsilon_1} \ e^{B(s)/4 \ t}
\end{eqnarray}
 with $G(t)=G_E(t)$ being the Sachs electric form factor relative to the first moment of GPDs and $A(t)$ relative to the second moment of GPDs.
  \begin{eqnarray}
G(t) =L_{1}^{4}/(L_{1}^2-t)^2 \ (4m_{p}^{2}- \mu \ t)/(4m_{p}^{2}- \ t); \ \
A(t)= L_{2}^4/(L_{2}^2-t)^2
 .\label{overlap}
 \end{eqnarray}
with the parameters:  $L^{2}_{1}=0.71 \ $GeV$^2$; $L_{2}^2=2 $~GeV$^2 $.
 $   \hat{s}=s \ e^{-i \pi/2}/s_{0} ;  \ \ \ s_{0}=1 \ {\rm GeV^2}$.
%
%
%

\label{sec:figures}
\begin{figure}[htb]
\includegraphics[width=0.32\textwidth] {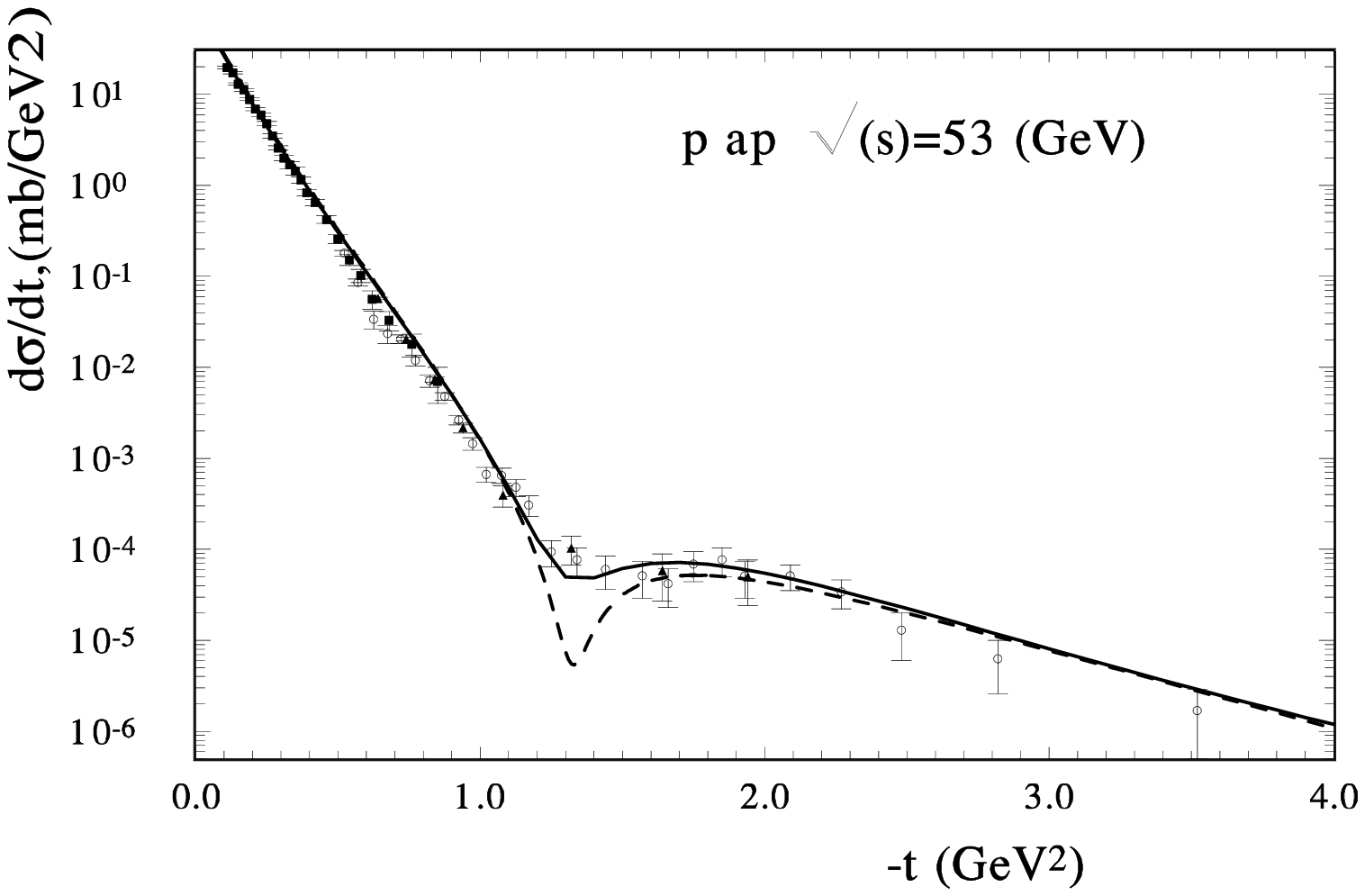}       
\includegraphics[width=0.32\textwidth] {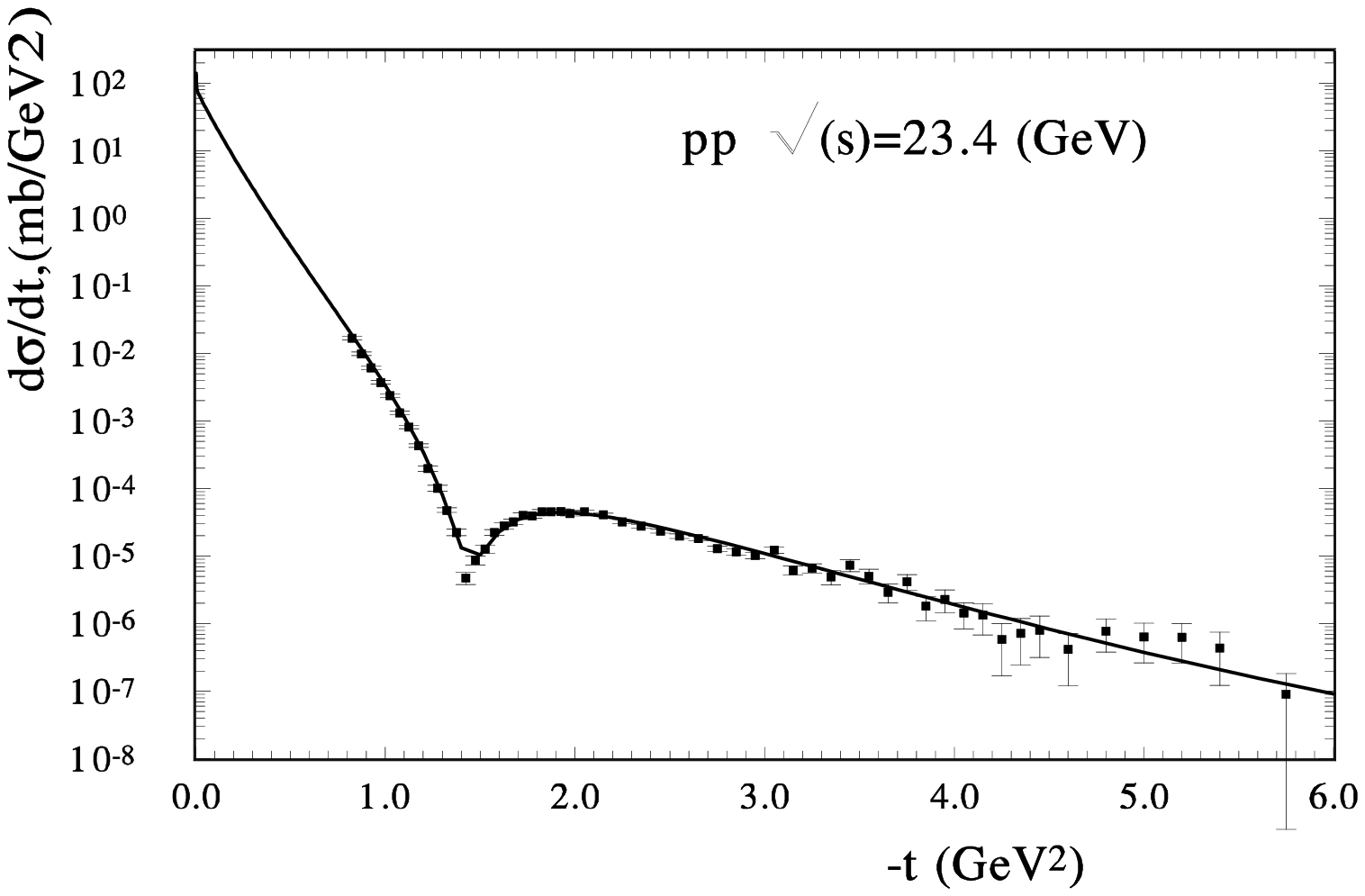}       
\includegraphics[width=0.32\textwidth] {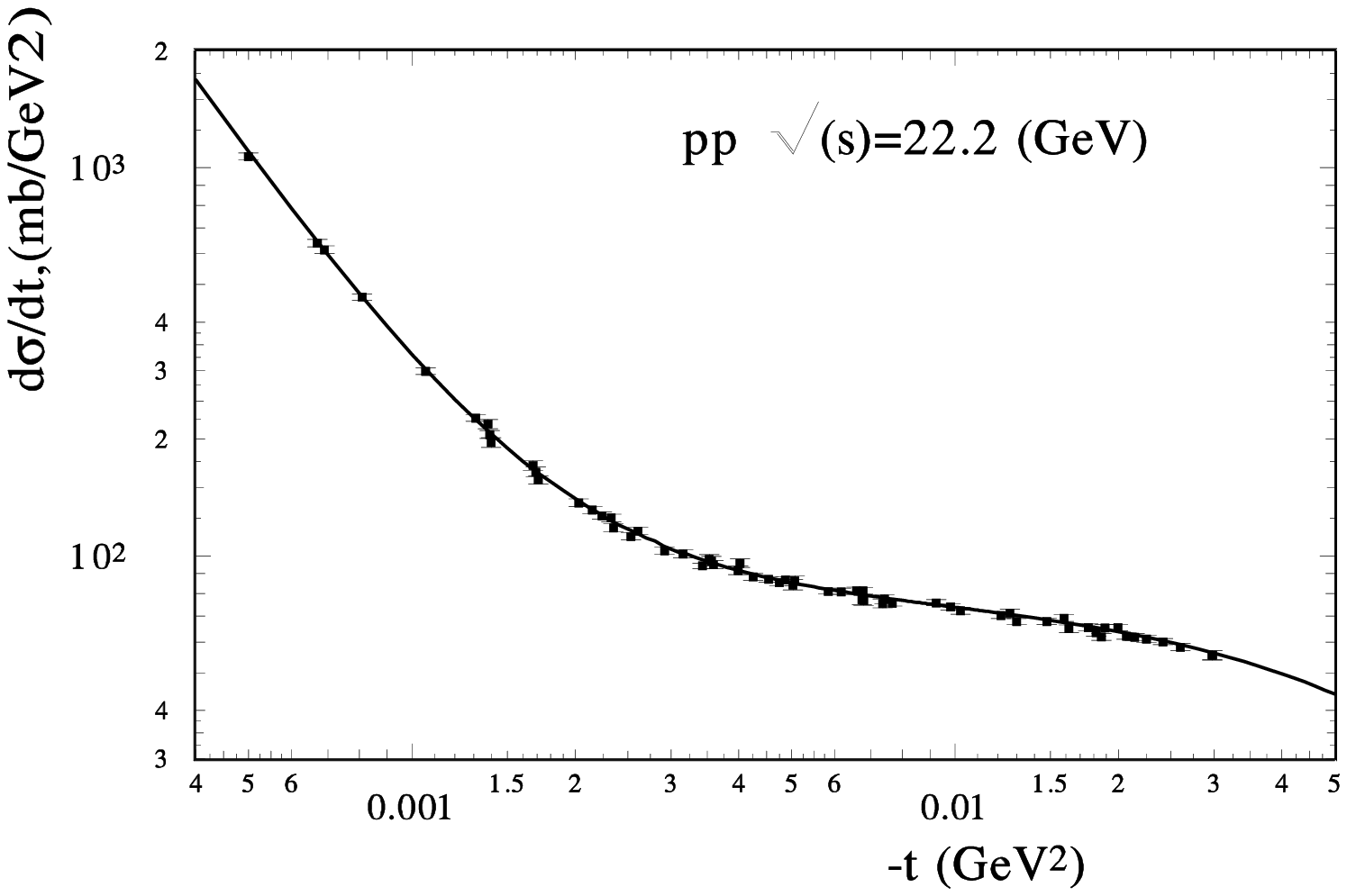}
\caption{
$d\sigma/dt$ are calculated in the model
for $p\bar{p} $ at $\sqrt{s}=52.8$ GeV (with [hard line] and without [dashed line] odderon contributions) and for
$pp$ at $\sqrt{s}=23.4$ and $\sqrt{s}=22.2$ GeV.
}\label{Fig:1}
\end{figure}

The final elastic  hadron scattering amplitude is obtained after unitarization of the  Born term.
    We have to calculate the eikonal phase
   and then obtain the final hadron scattering amplitude
   by integration of the eikonal of the scattering amplitude in the impact parameter representation.

\label{sec:figures}
\begin{figure}[htb]
\vspace{-1.5cm}
\includegraphics[width=0.5\textwidth] {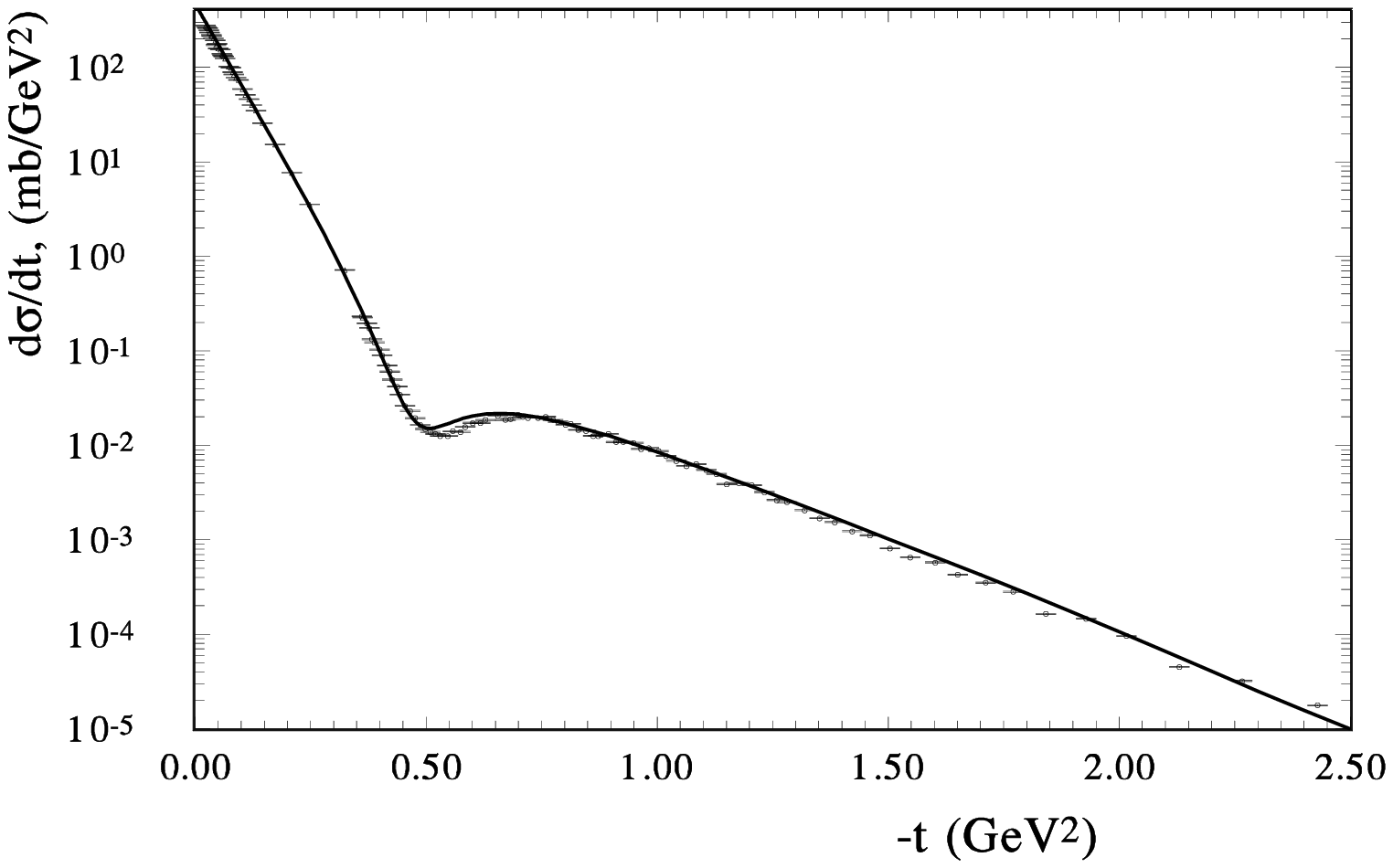}       
         \includegraphics[width=0.5\textwidth] {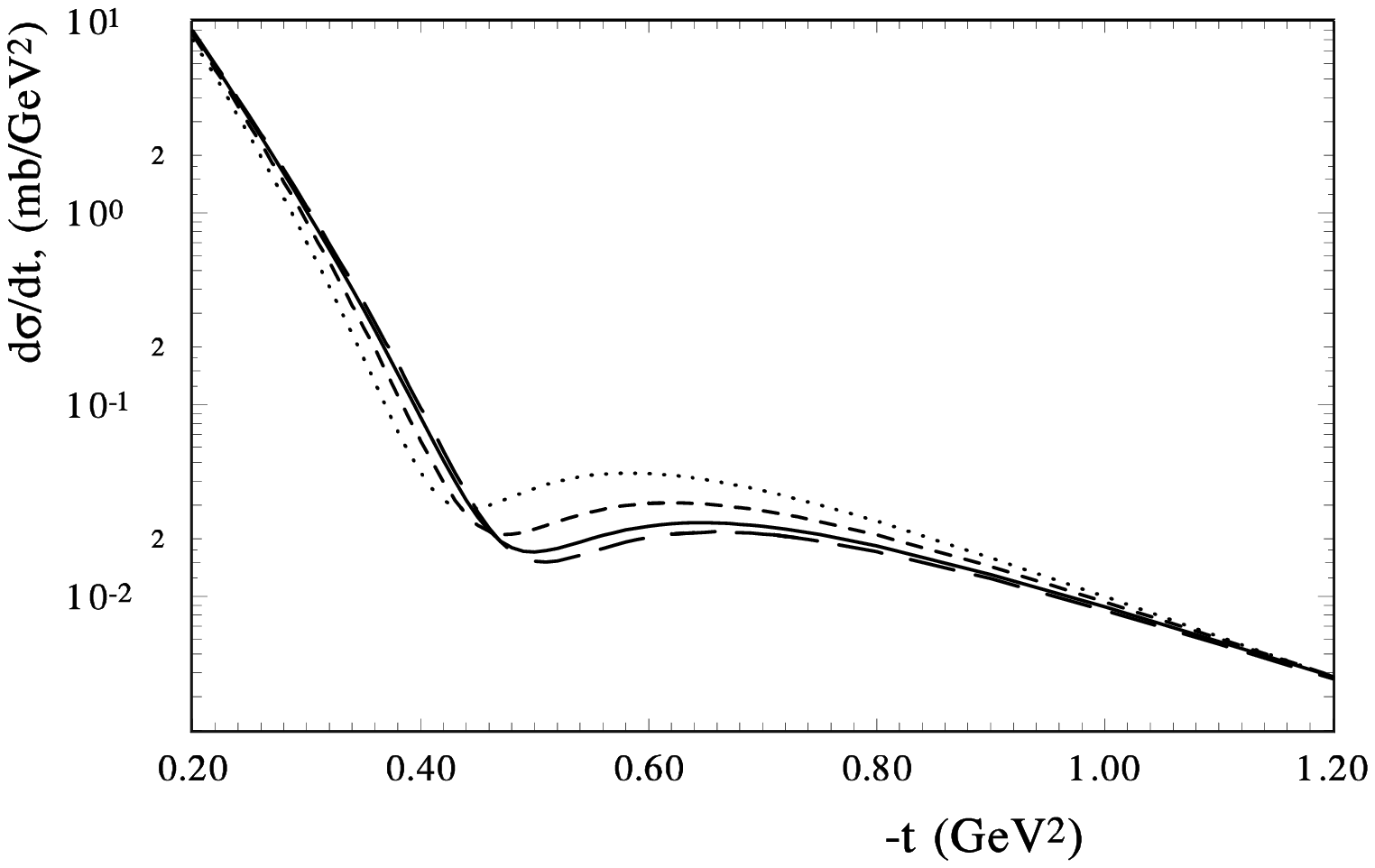}       
\caption{a)[left] Comparison of 
the model calculations with the experimental data  at $\sqrt{s}=7$  TeV; b)[right] $d\sigma/dt$ are calculated in
the model
(long-dashed, hard, short-dashed, and doted lines  at $\sqrt{s}=7, 8, 10 $ and $14$ TeV.
}\label{Fig:1}
\end{figure}


  The first (simplest) variant of the model  \cite{m1} has only $3$ high energy fitting parameters
 and $2$ low energy parameters, which reflect some small contribution
 coming from the different low energy terms.
   We check up the possible contribution of the hard pomeron   \cite{m1-hp}. We find that such contributios
   cannot see from the existing experimental data.
   Now in the slightly expanding variant of the model we take into account the odderon contributions.
   We assume that the odderon, as cross-odd 3 gluon state, has the vertex with the second form factor $A(t)$.
        As we take low energy data, we include the small contribution of the hadron spin-flip amplitude in the simplest form.
      $F_{sf}=h_{sf}*q*Exp[b_{sf}t]$.
  Now we  take all existing
      experimental data in the energy range $20 \leq \sqrt{s} \leq 7000 \ $GeV
      and the region of the momentum transfer $0.0007 \leq \ -t \ \leq 15 \ $GeV$^2$.

We do not include the data on the total cross sections
        $\sigma_{tot}(s)$  and $\rho(s)$.
       We also do not include the interpolated   and extrapolated data of Amaldi. 
%
%
As a result, one obtains $\sum \chi^2_i /N \simeq 1.5 $, where
the number of experimental points $N=2011 $.
The energy dependence of the scattering amplitude is determined
only by the single intercept and the logarithmic dependence on $s$ of the slope.
 Now we obtain a good description for $p\bar{p}$ scattering at  $\sqrt{s} = 52.8 $~GeV (Fig.~1(left panel)).
At this energy there are experimental data at small
(beginning at $-t=0.001 $~GeV $^2 $) and large (up to $-t=10 $~GeV $^2 $) momentum transfers.
The model reproduces both regions and provides a qualitative description of the dip region for all energies (Fig. 1b).

%
\begin{wraptable}{r}{0.45\textwidth}
\centerline{\begin{tabular}{|c|c|c|}
\hline
 $\sqrt{s}$, GeV & $\rho(t=0)$   & $\sigma_{tot}$, mb  \\\hline
  & &  \\
  22.2 & $0.0013 $        & $ 39.85 $   \\
   52.8 & $0.076 $        & $ 42.85 $   \\
   541 & $0.128 $        & $ 62.91  $   \\
   1800 & $0.127 $        & $ 76.25 $    \\
   7000 & $0.121$    &  $   95.9 $    \\
  8000 & $ 0.120$        &  $ 98.1  $     \\
 10000  & $ 0.119$       &  $101.6  $    \\
  14000  & $ 0.117 $    &  $ 107.3  $    \\
\hline
\end{tabular}}
\caption{The predictions  of the model. }
\label{tab:limits}
\end{wraptable}

    We present the new model of the hadron-hadron interaction at high energies.
  As we know, it is the only  model which describes all available high energy data
  in the Coulomb-hadron region and large momentum transfer.
   The model shows the contributions of the odderon with the same intercept as the pomeron. So it is the case of the
   maximal odderon.
   The energy dependence of the differential cross sections is determined  by only one intercept
 with $\epsilon =0.11$. 
 The real part of the
  hadron scattering amplitude is determined only by complex energy $s$ that satisfies the crossing-symmetries.
 The most important advantage of the model is that it is built on some  physical
     basis - two form factors which are calculated from GPDs.
     The model predictions for  $\sigma_{tot}$ and $\rho$ are shown in Table 1.
     They well coincide with the existing experimental data before the LHC era. \\


  {\bf Acknowledgments:}  I gratefully acknowledge the organizing committee and R. Orava for the invitation on the conference
  and the financial support. 
  This work is partially supported by WP8 of the hadron physics program of
the 8th EU program period.



\begin{footnotesize}

\end{footnotesize}
\end{document}